\documentclass[aps,pre,floats,floatfix,twocolumn]{revtex4}

\usepackage{ifthen}
\usepackage{ifpdf}
\usepackage{color}

\ifpdf
\usepackage{graphicx}
\usepackage{epstopdf}
\else
\usepackage{graphicx}
\usepackage{epsfig}
\fi

\usepackage{latexsym}
\usepackage{amsmath}
\usepackage{amssymb}
\usepackage{bm}
\usepackage{wasysym}

\newcommand{\eexp}{\mbox{e}^}

\newcommand{\amatrix}[1]{\begin{matrix} #1 \end{matrix}}

\newcommand{\mylabel}[1]{\label{#1}} 
\newcommand{\beq}{\begin{eqnarray}}
\newcommand{\eeq}{\end{eqnarray}} 
\newcommand{\be}[1]{\begin{eqnarray}\ifthenelse{#1=-1}
{\nonumber}{\ifthenelse{#1=0}{}{\mylabel{e#1}}}}
\newcommand{\ee}{\end{eqnarray}} 

\newcommand{\Eq}[1]{{Eq.~(\ref{#1})}} 
\newcommand{\Fig}[1]{{Fig.~\ref{#1}}} 
\newcommand{\hide}[1]{[hidden text]}
\renewcommand{\cite}[1]{\textcolor{blue}{[\onlinecite{#1}]}} 

\newcommand{\ei}{\hat{a}}
\newcommand{\eidag}{\hat{a}^{\dag}}
\newcommand{\hn}{\hat{n}}

\newcommand{\Jx}{\hat{J}_x}
\newcommand{\Jy}{\hat{J}_y}
\newcommand{\Jz}{\hat{J}_z}

\begin{document}

\title{Coherence dynamics of kicked Bose-Hubbard dimers: \\ Interferometric signatures of chaos}

\author{Christine Khripkov$^1$, Doron Cohen$^2$, and Amichay Vardi$^1$}
\affiliation{Departments of $^1$Chemistry and $^2$Physics, Ben-Gurion University of the Negev, Beer-Sheva 84105, Israel}

\begin{abstract}
We study the coherence dynamics of a  kicked two-mode Bose-Hubbard model starting with an arbitrary coherent spin preparation. For preparations in the chaotic regions of phase space we find a generic behavior with Floquet participation numbers that scale as the entire $N$-particle Hilbert space, leading to a rapid loss of single-particle coherence. However, the chaotic behavior is not uniform throughout the chaotic sea and unique statistics is found for preparations at the vicinity of hyperbolic points that are embedded in it. This is contrasted with the low $\log(N)$ participation that is responsible for the revivals in the vicinity of isolated hyperbolic instabilities.
\end{abstract}  


\maketitle

\section{Introduction}
\label{sec:intro}

One-particle coherence is the most distinct hallmark of Bose-Einstein condensation. The observation of matter-wave interference fringes served as  an unequivocal proof for the generation of dilute-gas Bose-Einstein condensates (BECs) \cite{Andrews97}  and the loss of their visibility was used as a sensitive probe for atom number squeezing \cite{Orzel01} and the superfluid to Mott-insulator quantum phase transition \cite{Greiner02} in BECs confined in periodic optical lattices. Coherent mean-field Josephson dynamics was demonstrated in double-well condensates \cite{Albiez05} and serves as a starting point for the construction of sub-shot-noise atom interferometers with Gaussian squeezed states \cite{Gross10,Riedel10}.

Complete many-body treatment of dilute-gas BECs with realistic particle numbers is normally beyond the scope of current theoretical methods. However, in certain instances it is possible to reduce computational complexity due to an effectively small number of contributing modes.  One particularly simple example is that of two coupled tightly bound BECs. Experimental examples are BECs in double-well potentials  \cite{Albiez05} and BECs of atoms with two optically coupled internal states (spinor BECs) \cite{Gross10,Riedel10}. Provided the interaction strength is sufficiently small with respect to the excitation energy of either condensate, such systems can be described by the tight-binding model of the two-mode Bose-Hubbard Hamiltonian (BHH) \cite{Leggett,Gati07}.  We note that deviations from the BHH model are obtained due to the participation of higher bands when the interaction strength becomes comparable with the excitation gap \cite{Sakmann09}.

Within the BHH model, it is particularly interesting to study the fluctuations of two-mode single-particle coherence \cite{Revivals,Boukobza09,Chuchem10}. Such fluctuations correspond to the collapses and revivals of multishot fringe visibility when the condensates are released and allowed to interfere. The natural initial conditions for such studies are coherent spin preparations \cite{Schumm05}, naturally obtained by manipulation of the superfluid ground state in the strong coupling regime. The observed coherence fluctuations are determined by the number~$M$ of participating eigenstates in the initial coherent preparation.   

In this work we study the one-particle coherence dynamics of two coupled BECs with initially coherent preparations. We assume the two-mode BHH model to be valid throughout the paper. In contrast to previous studies with time-independent parameters \cite{Revivals, Boukobza09,Chuchem10}, we consider here the case where the hopping term in the BHH is periodically modulated in time so as to produce a chaotic classical limit \cite{Strzys08}. This can be attained via modulation of the interwell barrier height in a double-well BEC realization or via the modulation of the coupling fields in the spinor BEC realization.
Specifically, we consider the realization of the kicked-top model \cite{Strzys08,Haake} where the the modulation is a sequence of kicks with period~$T$.
We contrast the previously studied integrable case ($T\rightarrow0$)  \cite{Vardi01,Boukobza09,Chuchem10,Khripkov12,Revivals,Zibold10} with the 
chaotic case (finite $T$), first introduced in the BHH context in Ref.~\cite{Strzys08}. 

The dynamics of the integrable $T\rightarrow 0$ model, like that of the Jaynes-Cummings model \cite{JC}, exhibits a series of collapses and revivals \cite{Revivals,Boukobza09,Chuchem10,Khripkov12}. These recurrences are manifest in the Rabi-Josephson population oscillations, as well as in the average fringe visibility, 
when the two condensates are released and allowed to interfere.  As shown in Refs.  \cite{Boukobza09,Chuchem10}, they  result from at most $M\sim \sqrt{N}$ (and for some preparations much fewer) participating eigenstates of the integrable Hamiltonian,  constituting any coherent preparation. 

For the nonintegrable model we find here that coherent states in the chaotic regions of phase space have far greater participation numbers of the order of the entire Hilbert space dimension ($M\sim N$). This leads to rapid loss of one-particle coherence and practically prevents the collapse and revival near elliptical 
or hyperbolic points. Furthermore, we observe different types of chaotic behavior: Coherent preparations located on hyperbolic points within the chaotic sea have a significantly lower participation number compared with those that reside inside the sea for which $M \approx N/2$. The latter agrees with random matrix theory (RMT), 
while the former can be regarded as arising from so-called scars \cite{scars,scars1,scars2}.

\section{Model}
\label{sec:model}

We consider a two-mode bosonic system that is described by the BHH. 
The hopping term is periodically modulated as a sequence of kicks, 
hence the BHH is formally identical to that of a kicked top \cite{Strzys08,Haake}:
\begin{equation}
\label{KBHH}
\mathcal{H} \ = \ U\Jz^2 
-\left[\sum_{n=-\infty}^{\infty}\delta\left(\frac{t}{T}-n\right)\right] K \Jx,
\end{equation}
where $\Jx{=}(\eidag_1 \ei_2{+}\eidag_2\ei_1)/2$,  $\Jy{=}(\eidag_1\ei_2{-}\eidag_2\ei_1)/2i$, and $\Jz{=}(\hn_1 {-} \hn_2)/2$ are defined in terms of annihilation and creation operators for a particle in mode $i=1,2$. Number conservation $\hn_1+ \hn_2=N$, with $\hn_i\equiv \eidag_i\ei_i$, implies angular momentum conservation with ${j=N/2}$. 
We note that the same dynamics can be realized by modulating the interaction strength $U$ via a Feshbach resonance, keeping the hopping term constant \cite{Strzys08}. Either way the evolution can be obtained by sequential applications of the Floquet operator 
\beq
\hat{\bm{F}}(K,U,T) \ \ = \ \ \exp(iTK\Jx) \ \exp(-iTU\Jz^2).
\eeq
Accordingly the evolution operator for an integer number of cycles is
\beq
\hat{\bm{U}}(t; K,U,T) \ \ = \ \ \Big[ \hat{\bm{F}}(K,U,T) \Big]^{t/T}. 
\eeq

The integrable $T\rightarrow 0$ limit of this expression is obtained by fixing~$t$ and taking ${n_{\text{steps}}=t/T \rightarrow\infty}$ using the Trotter product formula. This yields
\beq
\hat{\bm{U}}(t; K,U, T\rightarrow 0) =  \eexp{-it \mathcal{H}_0}, 
\ \ \ \ \ \ \ \mathcal{H}_0 = U\Jz^2 - K\Jx.
\eeq
The integrable Hamiltonian $\mathcal{H}_0$ features 
a single dimensionless interaction parameter \cite{Leggett,Gati07}
\beq
u \ \ = \ \ \frac{NU}{K} 
\eeq
that characterizes the dynamics. For $u{>}1$ a separatrix 
appears in its classical phase space \cite{Vardi01,Zibold10,Chuchem10} 
and \Eq{KBHH} becomes a variation of a pendulum, 
with the characteristic Josephson frequency 
\beq
\omega_J \ \  = \ \ \sqrt{(K+NU)K}  \ \ \approx \ \ (NUK)^{1/2}.
\eeq

If the period $T$ is finite, there is an additional dimensionless chaoticity parameter
\beq
K_{\text{Chirikov}} \ \ = \ \ T^2NUK \ \ =  \ \ (\omega_JT)^2.
\eeq
This corresponds to the standard definition of the dimensionless kicking-strength parameter in the Chirikov standard map \cite{rotor}. It is the ratio of two frequencies: the induced frequency $\omega_J$ and the driving frequency~$1/T$. If the former is much slower than the latter, the effect of the driving is adiabatic and hence the integrable limit is reached. Conversely, as $K_{\text{Chirikov}}$ is increased, the phase space dynamics follows the familiar route to chaos: Resonances become wider and eventually overlap, creating a connected chaotic sea.

\begin{figure}
\centering
\includegraphics[width=0.5\textwidth] {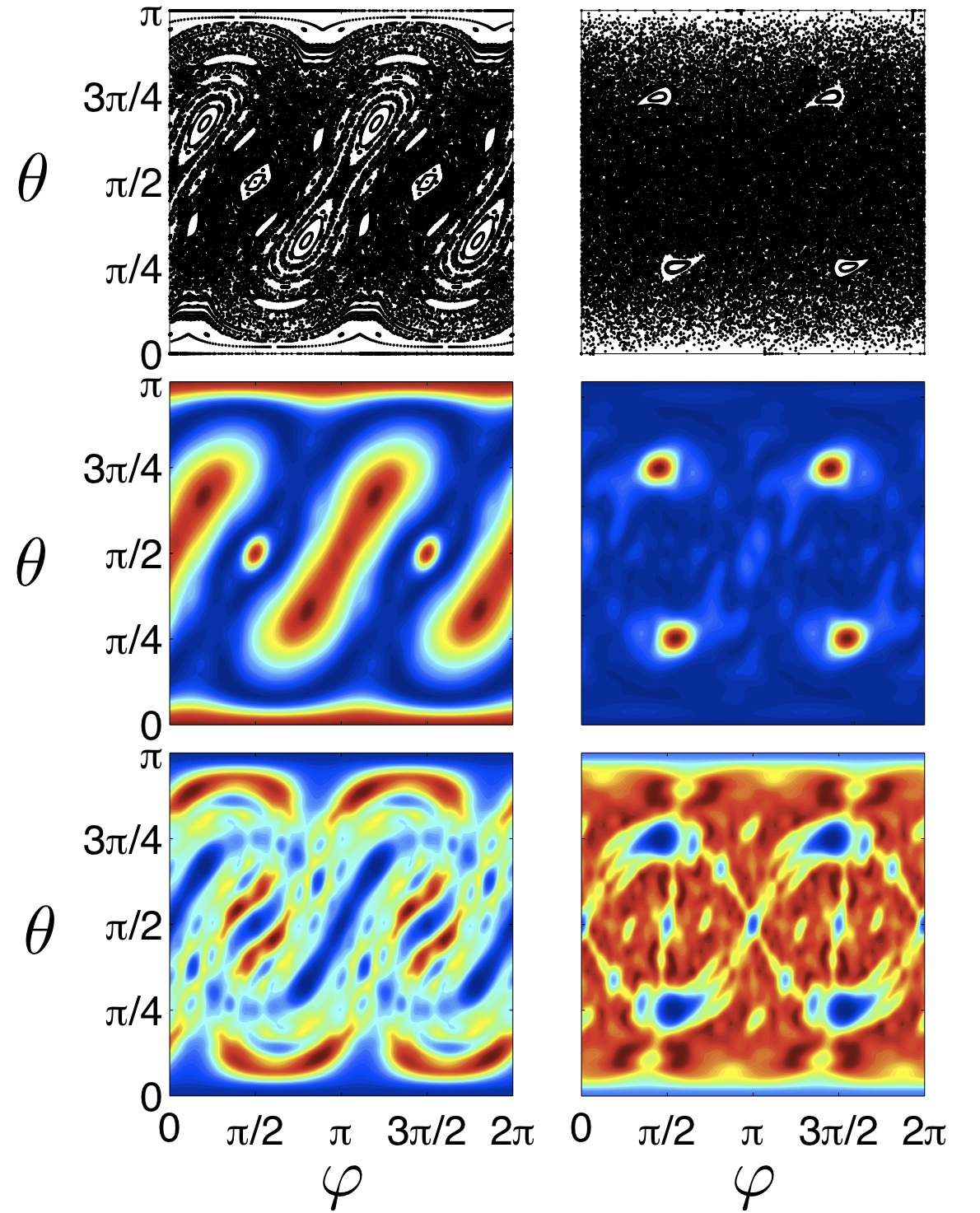}
\caption{(Color online) 
Classical phase space stroboscopic plots (top), 
time-averaged one-particle coherence $\bar{S}$ 
over $5000$ kicks (middle), 
and participation number $M$ (bottom)
for all spin coherent preparations $|\theta,\varphi\rangle$. 
The parameters here and later, unless specified otherwise, 
are $T=1$ and $K=\pi/2$. The number of particles is $N=200$, 
while $u=5/\pi$ (left) and $u=8/\pi$ (right). 
}
\label{fig1}
\end{figure}  

\section{Dynamics}
\label{sec:dynamics}

We study the quantum dynamics induced by the Hamiltonian \Eq{KBHH},  
starting from an arbitrary spin coherent state preparation
\begin{equation}
\label{SCS}
|\theta,\varphi\rangle \ \ \equiv \ \ \exp({-i\varphi\Jz})\exp({-i\theta\Jy}) \ |j,j\rangle.
\end{equation}
In these initial states all particles occupy a single superposition of the two modes, with a population imbalance $N\cos(\theta)$ and a relative phase~$\varphi$. Experimentally, such states can be prepared via a two-step process from the coherent ground state when $u\ll1$, in which $\theta$ is set by a coupling pulse 
and $\varphi$ by a bias pulse \cite{Zibold10}. 

The reduced one-particle density matrix of an $N$-particle quantum state 
is conveniently represented by the Bloch vector
\beq
{\bf S} \ \ \equiv \ \ \frac{2}{N} \left( \langle J_x \rangle, \langle J_y \rangle, \langle J_z \rangle  \right).
\eeq
The $S_z=\cos(\theta)$ component of the Bloch vector corresponds to the normalized population imbalance, while its azimuthal angle $\varphi=\arctan(S_y/S_x)$ corresponds to the relative phase between the modes. The expected fringe visibility if interferometry is carried out without further manipulation is 
\beq
g_{12}^{(1)} \ \ = \ \ \left(|S_x|^2+|S_y|^2\right)^{1/2},
\eeq
whereas the length $S=|\bm{S}|$ corresponds to the single-particle coherence:  
It is the best fringe visibility that one may expect to measure after proper manipulation, i.e., if it is allowed to perform any SU(2) rotation. Our coherent preparations \Eq{SCS} all have initial one-particle coherence value of $S=1$.

To the extent that an initial spin coherent state evolves only to other coherent states (without being deformed), the dynamics can be described by the mean-field equations, where the spin operators are replaced by $c$ numbers \cite{Vardi01} with ${\cal O}(1/N)$ accuracy and $S(t)=1$ for any time identically. Thus classicality in the $N$-body system is synonymous with one-particle coherence.
In the top row of \Fig{fig1} we show stroboscopic plots of the kicked-top classical dynamics for two representative values of the interaction parameter~$u$. 
The left panel depicts a mixed phase space with regular islands embedded in a chaotic sea, whereas in the right panel the islands shrink and the motion is chaotic almost throughout the entire phase space.  

While mean-field dynamics assumes perfect one-particle coherence, the nature of the classical motion is interlinked with its loss. In order to study the dynamics for the nonintegrable kicked BHH, we iterate the quantum state with the Floquet operator $\hat{\bm{F}}$ and calculate $S(t)$ over a time scale $t\gg\omega_J^{-1}$, where $t$ is the number of iterations ($T=1$). We then evaluate the long time average~$\bar{S}$ over $t=5000$ kicks, which is much larger than any semiclassical time scale, but still negligibly small with respect to the nonclassical time scales that are related to the tunnel coupling of separated phase space regions. We note parenthetically that these long tunneling times scale as $\sim\!\!\eexp{N}$ and are therefore unphysical in current experiments (e.g., $10^{13}\omega_J^{-1}$ for $N\approx100$) \cite{Khripkov12}. The value of~$\bar{S}$ for any initial coherent preparation is plotted in the middle row of \Fig{fig1}. From a comparison with the classical stroboscopic plots it is clear that one-particle coherence is lost completely 
for preparations lying in the chaotic regions of phase space. By contrast the coherence is better maintained in the regular island regions with near-unity values around elliptic fixed points.

\section{Fluctuations of an Integrable Dimer}
\label{sec:fluctuations}

We have previously studied the time evolution of $S(t)$ for the $T\rightarrow 0$ integrable dimer model \cite{Boukobza09,Chuchem10,Khripkov12}, starting with various coherent spin preparations that lead to different types of behavior. 
The key to the analysis lies in expanding the coherent states \Eq{SCS} as
\begin{equation}
|\theta,\varphi\rangle=\sum_\nu |\nu\rangle\langle\nu| \theta,\varphi\rangle
\label{ldosdef}
\end{equation} 
in the BHH eigenstates basis $|\nu\rangle$ pertinent to the specified interaction parameter $u$. 
The effective number of eigenstates that contribute to the wave-packet dynamics of \Eq{ldosdef} is evaluated by the participation number 
\beq
M \ \ = \ \ \left[ \sum_\nu p_\nu^2 \right]^{-1},
\eeq
where $p_\nu=|\langle\nu| \theta,\varphi\rangle|^2$. For example, in the strong interaction regime $u>N^2$, the eigenstates are merely Fock number states, so that $M\sim\sqrt{N}$, resulting in the collapse of coherence on a  time scale $(U\sqrt{N})^{-1}$ and its revival at $t=(UN)^{-1}$ \cite{Greiner02b}.  
The measurement of long time recurrences in this case has been utilized to experimentally detect effective higher-order interactions resulting from the dependence of $U$ on $N$ (see, e.g., Ref. \cite{Will10}).

The dynamics are far more complex in the Josephson interaction regime $1<u<N^2$
due to the coexistence of nearly linear and highly nonlinear phase space regions \cite{Boukobza09,Chuchem10,Khripkov12}. Consequently, by semiclassically evaluating the participation number we have identified a rich and nonuniversal dependence of $M$ on $N$ \cite{Chuchem10}. For example, the coherent preparations $|\pi/2,0\rangle$ and $|\pi/2,\pi\rangle$ are both characterized by equal population, but with very different participation numbers $M\approx\sqrt{u}$ and $M\approx\sqrt{u}\log(N/u)$, respectively. Consequently, one observes a substantial difference in their coherence dynamics \cite{Boukobza09}. 
These differences reflect the elliptic versus hyperbolic nature of the classical mean-field dynamics in the vicinity of the two points. Moreover, for other coherent preparations, with the same energy as that of $|\pi/2,\pi\rangle$, we find a much larger participation number $M \sim \sqrt{N}$ \cite{Chuchem10}. 

Summarizing these generic $M(N)$ dependencies for the integrable case, 
we have, depending on the characteristics of the classical motion,
\begin{equation}
\label{integrableM}
M\approx\left\{
\begin{array}{lr}
\sqrt{u}&{\rm elliptic~fixed~point}\\
\sqrt{u}\log(N/u)&{\rm hyperbolic~fixed~point}\\
\sqrt{N}\log(N/u)&{\rm separatrix~edge}.
\end{array}
\right.
\end{equation}
We reemphasize that a purely semiclassical picture, going beyond mean-field by accounting for the deformation of the initial Gaussian according to the classical motion, does not provide a satisfactory description of the dynamics. 
In order to predict the fluctuations it is essential to have a theory for~$M$.

\section{Scars}
\label{sec:scars}

In order to gain a similar understanding of the dynamics in the mixed phase space 
of the kicked-top Hamiltonian \Eq{KBHH}, we expand each coherent preparation \Eq{SCS} in the Floquet basis, meaning that $|\nu\rangle$ are now redefined as the eigenstates of the Floquet propagator $\hat F$. The participation number $M$ is then evaluated using the expansion coefficients. The results for all coherent preparations are shown in the bottom row of \Fig{fig1} for two representative parameter values. As expected, the participation is generally far greater for preparations corresponding to classical points within the chaotic sea compared with preparations in regular regions. 
However, not all the points within the chaotic sea have the same participation number. As seen in the bottom right panel of \Fig{fig1}, there are evident scars in the vicinity of the hyperbolic fixed points, though they are immersed in a chaotic sea.

\begin{figure}
\centering
\includegraphics[width=0.9\hsize] {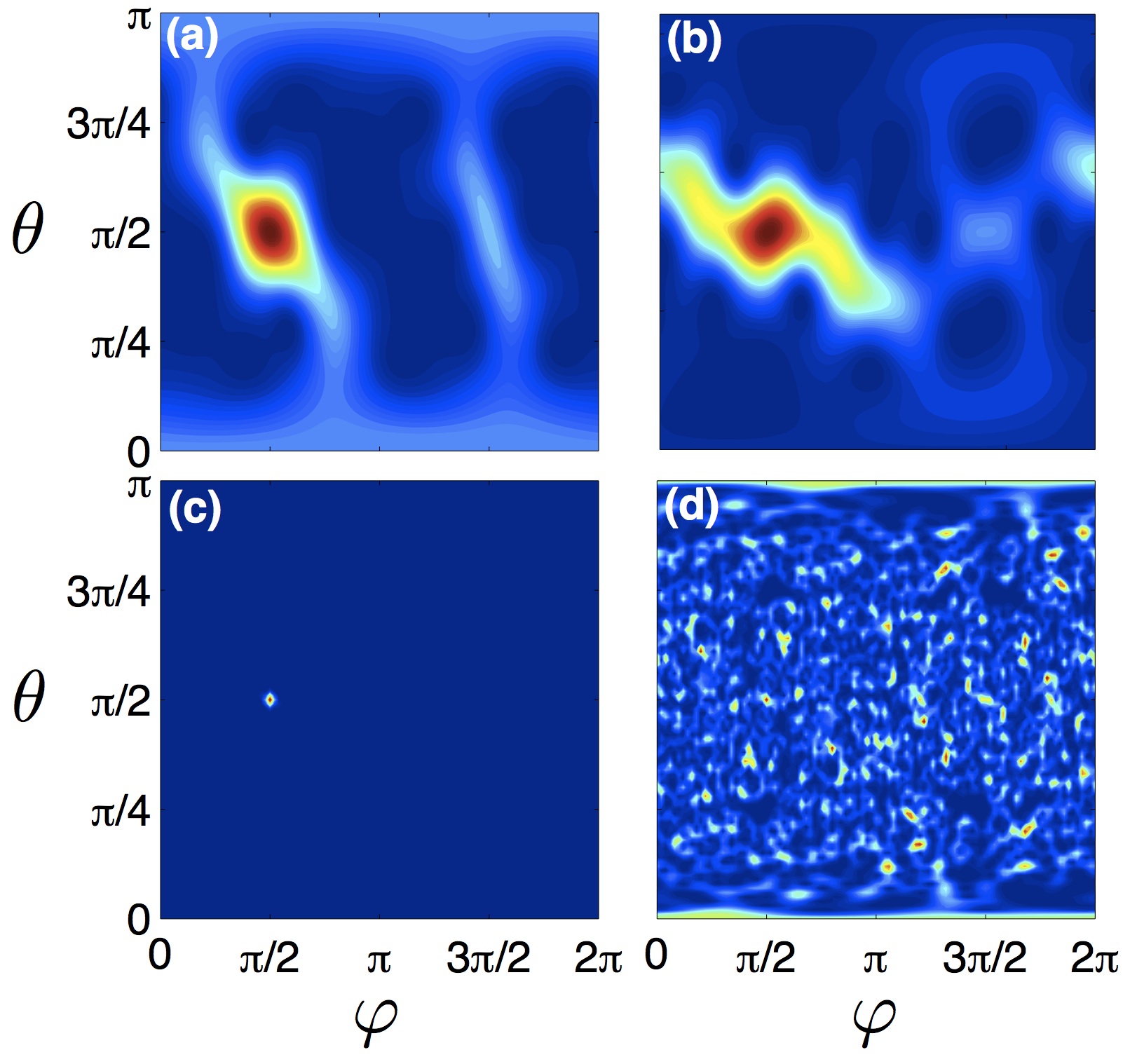}
\caption{(Color online) 
Husimi distributions after $t=1000$ iterations, 
starting with the spin coherent state $|\pi/2,\pi/2\rangle$.
The interaction is (a) and (c) $u=5/\pi$ and (b) and (d) $u=8/\pi$,
with (a) and (b) $N=20$ and (c) and (d) $N=2000$.
}
\label{fig2}
\end{figure}  

\begin{figure}
\centering
\includegraphics[width=0.5\textwidth] {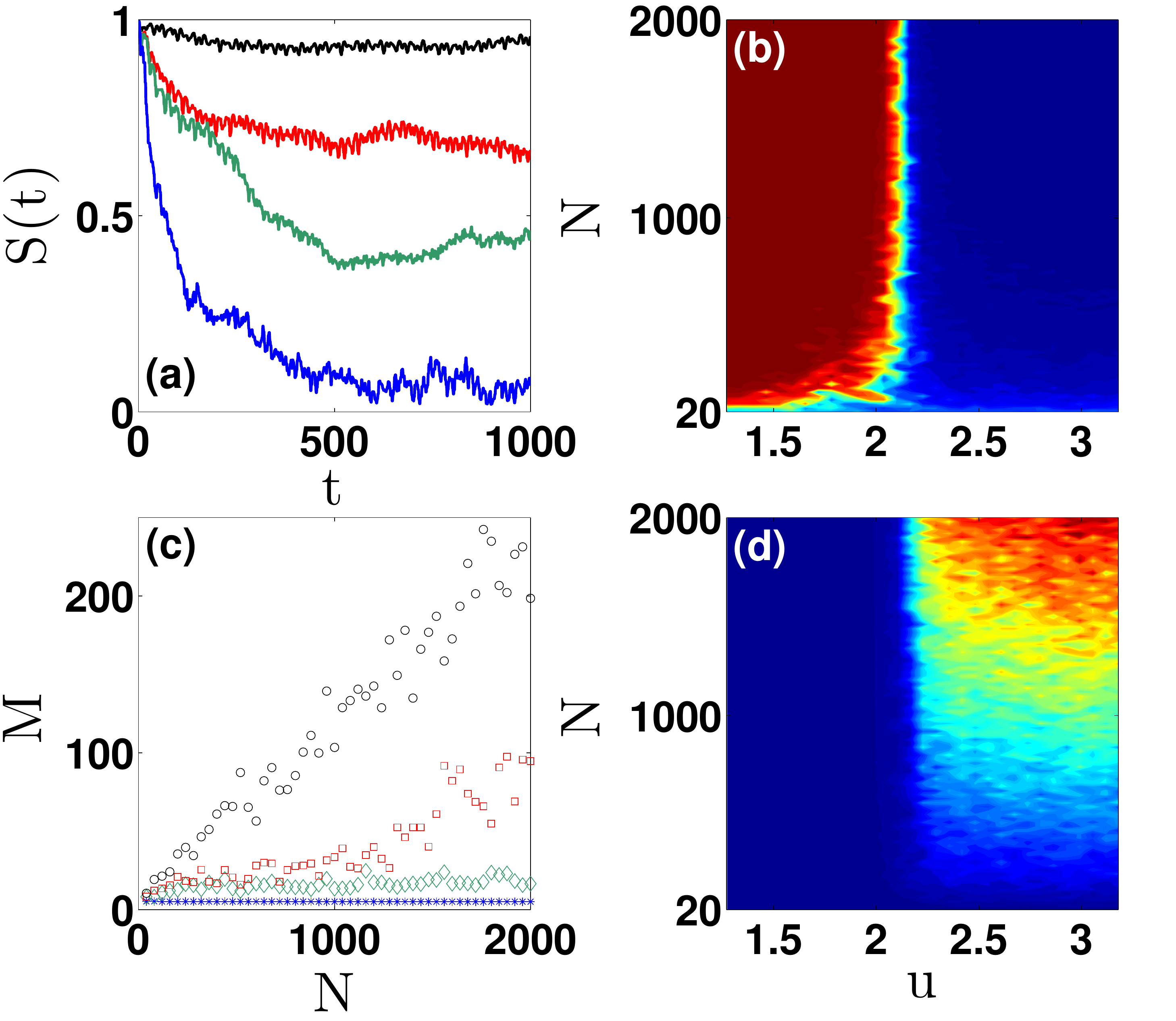}
\caption{(Color online) 
Participation number $M$ and time evolution of the coherence $S(t)$
for the initial preparation $|\pi/2,\pi/2\rangle$. (a) From top to bottom, $S(t)$ for $u=6.4/\pi$, $6.64/\pi$, $6.66/\pi$, and $6.8/\pi$, with $N=1200$.  
(b) Time-averaged coherence $\bar S$ as a function of $u$ and $N$.
(c) Bottom to top, the participation $M$ as a function of $N$ for $u=4.8/\pi$, $6.6/\pi$, $6.8/\pi$, and $8.8/\pi$.   
(d) Participation $M$ as a function of $u$ and $N$.
}
\label{fig3}
\end{figure}  

\section{Transition to chaos}
\label{sec:transition}

We turn our attention to the coherent preparation $|\pi/2,\pi/2\rangle$. 
As shown in \Fig{fig1}, it resides on a fixed point of the classical dynamics 
that undergoes a transition from being elliptical to hyperbolic as $u$ is increased. 
Linearizing the kicked-top map around this fixed point (see the Appendix), 
we obtain the Lyapunov instability exponent
\beq
&& \lambda \ = \ \frac{1}{T_0}\log\left(b+\sqrt{b^2-1}\right), 
\label{lyapunov}
\\ \nonumber
&& b \ \equiv \ \left(\frac{\pi u}{4}\right)^2\left[1{-}\cos(\pi u)\right] +\left(\frac{\pi u}{2}\right)\sin(\pi u) + \cos(\pi u),
\eeq
where $T_0=4$ is the fixed point period, i.e., the number of kicks required to cycle it to its initial position. 
From \Eq{lyapunov} it is clear that the fixed point becomes hyperbolic at $u\approx 2.1$. 
Note that for larger values of $u$ there are narrow windows around ${u=4,6,8,\dots}$, 
corresponding to $2\pi n$ kicks, within which elliptic behavior is restored.

The classical transition from elliptic to hyperbolic dynamics is reflected in the quantum simulations. In \Fig{fig2} we  start with the spin coherent state $|\pi/2,\pi/2\rangle$ and plot the Husimi distribution $\langle\theta,\varphi|{\hat\rho(t)}|\theta,\varphi\rangle$,  where $\hat \rho(t)$ is the $N$-particle density matrix after $t=1000$ cycles. We contrast the interaction parameter values $u=5/\pi$ (elliptic, left) and $u=8/\pi$ (hyperbolic, embedded in a chaotic sea, right) for two representative values of $N$. Remembering that $1/N$ serves as an effective Planck constant for the many-body system, one expects a quantum-to-classical correspondence at large $N$ when the size of the preparation is small compared with the Planck cell.  
Thus, as shown in Figs.~\ref{fig2}(c) and \ref{fig2}(d), for sufficiently large $N$ there is a transition from conservation of coherence over long times 
(when $u$ is such that the fixed point is elliptic [\Fig{fig2}(c)]) to a rapid loss of coherence due to the smearing of the distribution throughout the chaotic sea (when $u$ is such that the fixed point is hyperbolic [\Fig{fig2}(d)]). The difference between the two cases is blurred for small particle numbers when the Planck cell becomes larger than the initial preparation [Figs.~\ref{fig2}(a) and \ref{fig2}(b)] and quantum-classical correspondence is lost.

The resulting one-particle coherence dynamics is shown in \Fig{fig3}(a). In agreement with the above discussion, coherence is maintained for values of $u$ 
below the elliptic-to-hyperbolic transition and it is abruptly lost above it. The time-averaged coherence $\bar{S}$ is plotted throughout the $(u,N)$ parameter space 
in \Fig{fig3}(b), highlighting the sharp change at the classical transition and the loss of quantum-classical correspondence at low $N$.

A similar sharp transition is observed for the participation number $M$ of this preparation, as seen in Figs.~\ref{fig3}(c) and \ref{fig3}(d). For small $u$ the fixed point is elliptic, and hence $M$ has no dependence on $N$, as implied by \Eq{integrableM}. However, with the onset of chaos, when $u$ is increased, 
we obtain a different linear dependence of $M$ on $N$. Recall that $N+1$ with $N=2j$ is the dimension of the many-body Hilbert space, hence we have here an agreement with the semiclassical quantum-ergodic picture: A minimal Gaussian representing the initial coherent state has a nonzero overlap with all the eigenstates that reside in the chaotic sea.

\begin{figure}
\centering
\includegraphics[width=0.5\textwidth] {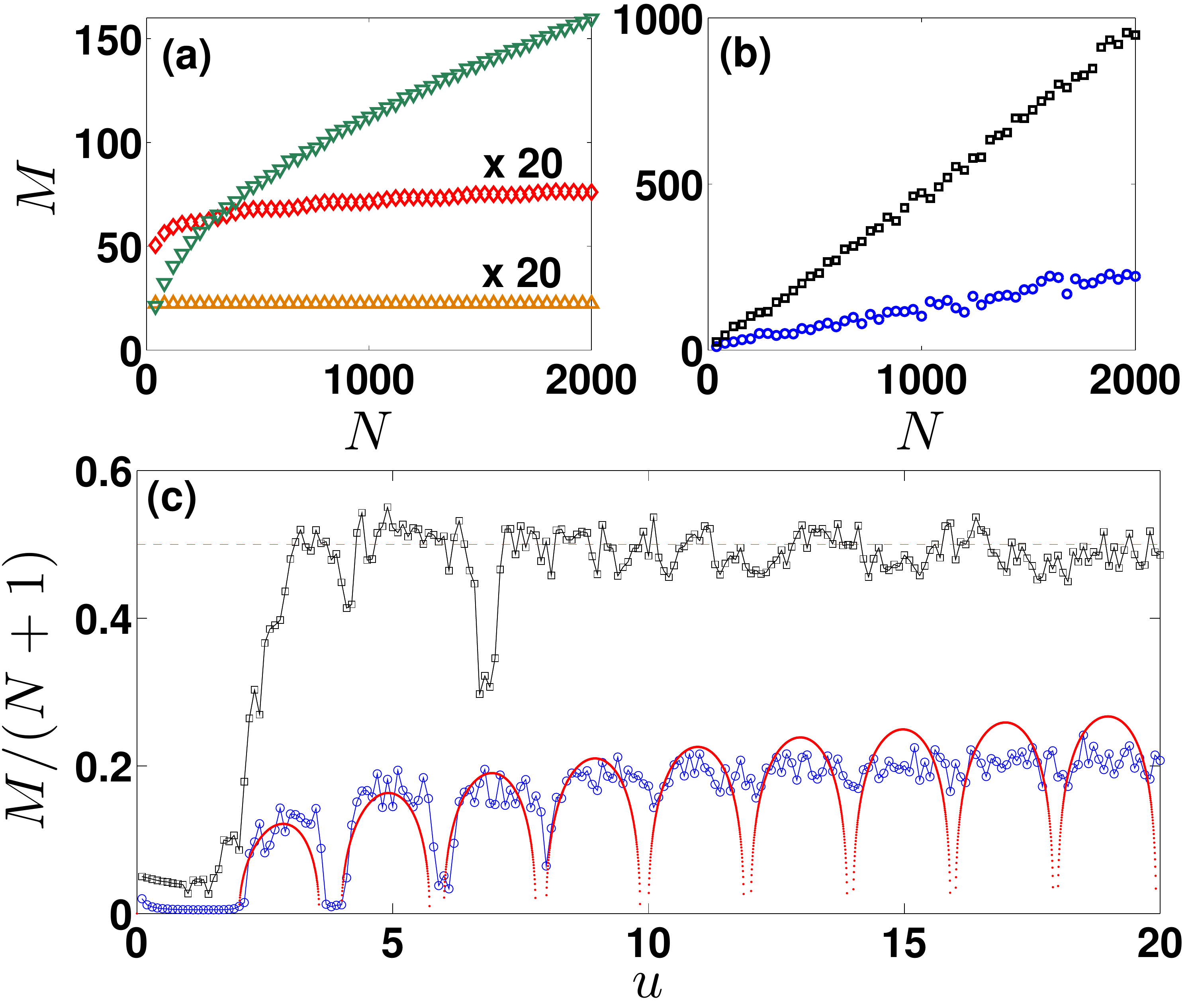}
\caption{(Color online) 
(a) Dependence of $M$ on $N$ for three preparations 
in the case of the integrable model:
at elliptic point $|\pi/2,0\rangle$ (bottom $\triangle$), 
at hyperbolic point $|\pi/2,\pi\rangle$ (middle $\diamond$), 
and on the separatrix edge $|0,0.9\pi \rangle$ (top $\triangledown$).
In (b) we contrast the results that are obtained 
in the case of the kicked system for preparations 
that reside on the hyperbolic point $|\pi/2,\pi/2\rangle$ (blue $\circ$) 
and on a nearby chaotic point $|\pi/3,\pi/3\rangle$ (black $\square$). For both (a) and (b) $u=9/\pi$.
(c) Dependence of $M$ on $u$ in the two cases of (b) with ${N=1000}$. The dashed and dotted red lines depict the RMT estimation $M\approx N/2$ and the semiclassical prediction of \Eq{hyperM}, respectively.  
}
\label{fig4}
\end{figure}  

\section{Wave-function statistics} 
\label{sec:statistics}

Figure \ref{fig4} summarizes the entire range of participation number behavior 
studied so far for the various initial coherent preparations. Figure \ref{fig4}(a) includes the previously studied cases for the integrable BHH model, whereas \Fig{fig4}(b) shows the linear scaling that we observe in the chaotic case. The large $M$ that characterizes chaotic sea preparations is responsible for the observed rapid loss of the one-particle coherence in \Fig{fig3}(a). 

The observed scars in \Fig{fig1} are characterized by a significantly smaller $M/N$ ratio. In Figs. \ref{fig4}(b) and \ref{fig4}(c) we contrast the $M/N$ dependence for a preparation at the hyperbolic point with that associated with preparation at a nearby fully chaotic point. The former is scar affected, while the latter exhibits the expected RMT dependence $M\approx N/2$ \cite{GUE}. In the vicinity of the hyperbolic point we expect this result to be semiclassically suppressed \cite{scars1,scars2} by a factor 
\beq
F_{\text{SC}} \ \ = \ \ \sum_{s=-\infty}^{\infty} \frac{1}{\cosh(\lambda s)}.
\eeq
Together with the RMT statistical factor $F_{\text{RMT}}=2$, which is implied by the Gaussian unitary ensemble, it gives the overall suppression and hence 
\beq
\label{hyperM}
M =  \frac{N}{ F_{\text{RMT}} \ F_{\text{SC}} } 
\approx 
\left[ \sum_{s=-\infty}^{\infty} \frac{1}{\cosh(\lambda s)} \right]^{-1} \frac{N}{2}.
\eeq
Substituting the Lyapunov instability exponent $\lambda$ from \Eq{lyapunov}
with no adjustable  parameters, we obtain very good agreement between this estimate and the hyperbolic point participation ratio [see \Fig{fig4}(c)]. The plot reflects the fixed-point transition from elliptical to hyperbolic behavior, with narrow windows of restored ellipticity around ${u=4,6,8,...}$, as discussed in section \ref{sec:transition}.

\begin{figure}
\centering
\vspace{5mm}
\includegraphics[width=0.5\textwidth] {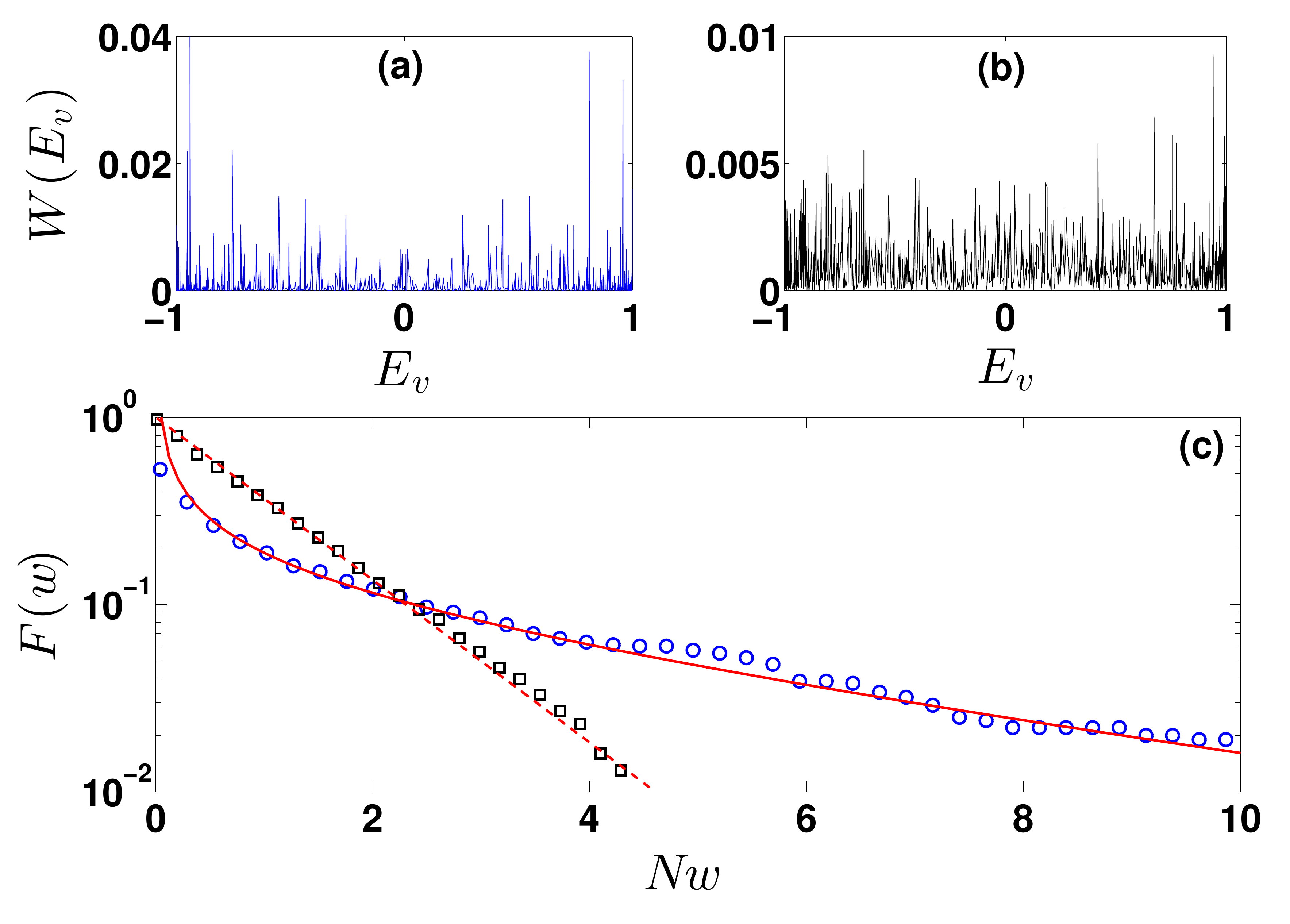}
\caption{(Color online) 
Wave-function statistics for the two representative 
chaotic sea preparations: 
The local density of states is plotted for (a) $|\pi/2,\pi/2\rangle$ and (b)
 $|\pi/3,\pi/3\rangle$. Note the different vertical scale.
(c) Comparison of the intensity statistics of (a) (blue, $\circ$) and (b) (black, $\square$). The dashed and solid lines are based on the RMT and the semiclassical theories (see the text). The parameters are $N=1000$ and $u=9/\pi$.  
}
\label{fig5}
\end{figure}  

The local density of states $W(E_\nu)\equiv|\langle \nu | \theta,\varphi\rangle|^2$ provides more detailed statistical information on the participation of the eigenstates 
in a given coherent preparation. Here $\exp(-i2\pi E_{\nu})$ are the eigenvalues of the Floquet operator $\hat{\bm{F}}$. Figure \ref{fig5} compares $W(E_\nu)$ for the two coherent preparations: hyperbolic versus generic chaotic points. One observes that the state $|\pi/2,\pi/2\rangle$ projects preferably onto a subset of eigenstates, reflecting that the latter have a significantly larger weight at the fixed point (scarring).  

To better quantify the eigenvector statistics of the two representative preparations, 
we plot the inverse cumulative histogram: $F(w)$ is the fraction of eigenstates  
with intensity $W(E_{\nu})>w$. As expected, the RMT statistics follows the Porter- Thomas law
\beq
F(x) \ \ = \ \ \exp(-x), 
\eeq
where $x=Nw$. In contrast, the statistics in the vicinity of the hyperbolic point 
follow a significantly different functional dependence \cite{scars1,scars2}, 
namely, 
\beq
F(x) \ \ \propto \ \ x^{-1/2}\exp(-\gamma x),
\eeq
where $\gamma$ is a fitting parameter that is proportional to the instability exponent $\lambda$.

\section{Summary}

Considering the coherence dynamics of a nonintegrable kicked-top BHH, 
one observes that an initial spin coherent preparation residing in 
the chaotic regions of the mixed phase space contains an $\mathcal{O}(1)$ fraction 
of the quantum eigenstates. This leads to an abrupt irrecoverable loss 
of single-particle coherence, as opposed to the collapse and revival dynamics 
that is obtained for the integrable model. Within the chaotic sea we find  
two distinct types of linear dependence reflecting different wave-function 
statistics. Namely, the lower participation number in the case of a wave-packet that 
is launched at a hyperbolic point reflects the wide distribution of 
overlaps due to scarring. 

The experimental consequence for coupled BECs is a rich variety of fringe visibility dynamics depending on the population imbalance and relative phase of the initial coherent preparation. In fact, using different preparations as in Ref.~\cite{Zibold10} and monitoring the visibility of interference fringes over time, it is possible to obtain a tomographic scan of the mixed phase space with chaotic regions leading to rapid loss of fringe visibility, as opposed to long-time coherence in regular island regions. Moreover, the loss of fringe visibility can be used for the detection of scars within the chaotic sea. 


\section*{ \MakeUppercase{Acknowledgments}}
We thank Lev Kaplan for a useful communication.
This research was supported by the Israel Science Foundation (Grants No. 346/11 and No. 29/11) and by the U.S.-Israel Binational Science Foundation.


\onecolumngrid

\section*{ \MakeUppercase{Appendix: Details of calculations}}

Consider the kicked-top map with $T=1$, $K=\pi/2$, and $U=w/2$ such that $u=(2/\pi)w$. Using the notation ${\bm{S}=(X,Y,Z)}$, the classical map is  
\beq
X' &=& X \cos(wZ) - Y  \sin(wZ), \\
Y' &=& Z, \\
Z' &=& -X \sin(wZ) - Y  \cos(wZ). 
\eeq
The fixed point $|\pi/2,\pi\rangle$ is cycled from ${\bm{S}=(0,1,0)}$ to $(0,0,-1)$ to $(0,-1,0)$ to $(0,0,1)$ repeatedly and hence has period $T_0=4$. The linearized transformation of $(X,Z)$ involves the matrix
\beq
\bm{M}=\left(\amatrix{
\cos(2w) + w\cos(w)\sin(w) & 
-2w\cos(w)^2 - w^2\cos(w)\sin(w) + w\sin(w)^2 + \sin(2 w) \\
-\sin(w) (2 \cos(w) + w\sin(w) & 
\cos(w)^2 + 3 w\cos(w) \sin(w) + (w^2-1) \sin(w)^2
}\right).
\eeq
The Lyapunov instability exponents $\pm\lambda$ equal the logarithm of eigenvalues of this matrix divided by $T_0$.   

\twocolumngrid


\clearpage
\end{document}